# STCTM: a forward modeling and retrieval framework for stellar contamination and stellar spectra

Caroline Piaulet-Ghorayeb[1, 2]

**1** Department of Astronomy & Astrophysics, University of Chicago, 5640 South Ellis Avenue, Chicago, IL 60637, USA **2** E. Margaret Burbidge Prize Postdoctoral Fellow

## Summary

Transmission spectroscopy is a key avenue for the near-term study of small-planet atmospheres and the most promising method when it comes to searching for atmospheres on temperate rocky worlds, which are often too cold for planetary emission to be detectable. At the same time, the small planets that are most amenable for such atmospheric probes orbit small M dwarf stars. This "M-dwarf opportunity" has encountered a major challenge because of late-type stars' magnetic activity, which lead to the formation of spots and faculae at their surface. If inhomogeneously distributed throughout the photosphere, this phenomenon can give rise to "stellar contamination," or the transit light source effect (TLSE). Specifically, the TLSE describes the fact that spectral contrasts between bright and dark spots at the stellar surface outside of the transit chord can leave wavelength-dependent imprints in transmission spectra that may be mistaken for planetary atmosphere absorption.

As the field becomes increasingly ambitious in the search for signs of even thin atmospheres on small exoplanets, the TLSE is becoming a limiting factor, and it becomes imperative to develop robust inference methods to disentangle planetary and stellar contributions to the observed spectra. Here, I present `stctm`, the STellar ConTamination Modeling framework, a flexible Bayesian retrieval framework to model the impact of the TLSE on any exoplanet transmission spectrum, and infer the range of stellar surface parameters that are compatible with the observations in the absence of any planetary contribution. With the `exotune` sub-module, users can also perform retrievals directly on out-of-transit stellar spectra in order to place data-driven priors on the extent to which the TLSE can impact any planet's transmission spectrum. The input data formats, stellar models, and fitted parameters are easily tunable using human-readable files and the code is fully parallelized to enable fast inferences.

## Statement of need

The interpretation of high-precision exoplanet transmission spectra from facilities such as the Hubble Space Telescope (HST) and the James Webb Space Telescope (JWST) is increasingly dependent on a robust accounting for the effects of stellar contamination, particularly for small planets orbiting small stars. Despite a growing awareness of the Transit Light Source Effect (TLSE; (Rackham et al., 2018; TRAPPIST-1 JWST Community Initiative et al., 2024)), the community currently lacks flexible, open-source tools that allow for robust modeling and retrieval of stellar contamination signatures. Further, uncertainties in stellar models motivate flexible implementations with reproducible model setups supporting any user-specified stellar model source, such as PHOENIX or SPHINX model grids (Husser et al., 2013; Iyer et al., 2023).

While some forward models have been developed to simulate the impact of stellar heterogeneity on transmission spectra, these tools are either not publicly available, computationally intractable due to their serial-mode-only implementation, part of much larger codes that require more advanced user training (e.g. atmospheric retrievals), or not designed for inference. Further, the community lacks frameworks that enable to retrieve stellar surface properties from both observed planetary transmission spectra and out-of-transit stellar spectra in a Bayesian context. This gap limits our ability to quantify uncertainties in exoplanet atmospheric properties and to test the robustness of atmospheric detections.

`stctm` addresses this need by providing an open-source, modular, and user-friendly framework. It allows users to model a wide range of stellar surface configurations leveraging any spectral models, and to infer which stellar parameters could explain observations without invoking planetary absorption. It also supports retrievals on out-of-transit stellar spectra to independently assess the extent of potential stellar contamination by the host star. By enabling flexible, fast, and reproducible inference of the TLSE, `stctm` empowers the community to critically assess the reliability of exoplanet atmosphere detections.

## Main features of the code

The user inputs are communicated to the code via an input `.toml` file for both TLSE retrievals and inferences from out-of-transit stellar spectra. The code follows similar phases for both types of retrievals:

- Reading in and parsing of the inputs (data file, stellar models file, saving options, MCMC fit setup)
- Running the MCMC fit
- Post-processing to create diagnostic plots, record model comparison and goodness-of-fit metrics, produce publication-ready figures, and store sample spectra and parameters for post-processing and publication support and reproducibility (e.g. Zenodo)

`exotune` retrievals on out-of-transit stellar spectra have an additional (optional) pre-processing step, allowing users to:

- start from a full time-series of spectra as the input (e.g. the output from Stage 3 of the `Eureka!`(Bell et al., 2022) pipeline which is widely used in the community), rather than simply a pre-computed stellar spectrum
- exclude certain time intervals/exposures (e.g. containing the transit) and wavelength ranges (e.g. saturated regions) when computing the median spectrum to perform the retrieval over

For `exotune` retrievals, an error inflation parameter can be leveraged to account for the often-large mismatch between the data and stellar models.

## Documentation

The full documentation for `stctm` with installation, testing instructions, and real-data example retrievals on transmission spectra and out-of-transit stellar spectra are available at https://stctm.readthedocs.io/. A description of `stctm` can be found in several of the early papers that employed it (see next section).

## Uses of STCTM in the literature

`stctm` has been applied widely to the interpretation of transmission spectra of rocky planets and small sub-Neptunes, including in (Ahrer et al., 2025; Lim et al., 2023; Piaulet-Ghorayeb et al., 2025, 2024; Radica et al., 2025; Roy et al., 2023).

## Future Developments

The latest version of `stctm` at the time of writing (v2.1.1) supports MCMC retrievals on transmission spectra and on out-of-transit stellar spectra (`exotune`), and provides model comparison statistics, model and parameter samples as well as publication-ready figures. Future versions will expand on these functionalities to include user-friendly scripts tailored to post-processing only for an already-run retrieval (creating custom plots), as well as a Nested Sampling alternative for the retrievals. Users are encouraged to propose or contribute any other features.

## Similar Tools

Here are a few open-source codes that offer functionalities focused on retrievals of the TLSE or on out-of-transit stellar spectra:

- Generic atmospheric retrievals (including TLSE-only retrievals on transmission spectra): `POSEIDON` (MacDonald & Madhusudhan, 2024)
- Retrievals on out-of-transit stellar spectra (Nested Sampling, serial run mode only): `StellarFit` (Radica et al., 2025)

## Acknowledgements

CPG thanks Warrick Ball for pre-review suggestions that improved the code documentation and ease of set-up for the user.`stctm` relies heavily on other Python libraries which include `numpy` (Harris et al., 2020), `scipy` (Virtanen et al., 2020), `astropy` (Astropy Collaboration et al., 2018, 2013), `matplotlib` (Hunter, 2007), `pandas`(team, 2020), `emcee`(Foreman-Mackey et al., 2013), `corner` (Foreman-Mackey, 2016), and `pysynphot` (Horne, 2013). Users are also strongly encouraged to use `msg`(Townsend & Lopez, 2023) to obtain the grids of stellar models used in the inference step.

CPG also acknowledges support from the E. Margaret Burbidge Prize Postdoctoral Fellowship from the Brinson Foundation. She thanks R. MacDonald, O. Lim, and M. Radica for helpful conversations that helped shape `stctm`.